\documentclass[journal=jacsat,manuscript=article]{achemso}

\usepackage[version=3]{mhchem} 
\usepackage[T1]{fontenc}
\usepackage{multirow}

\author{Kashinath T. Chavan}
\email{ktchavan99@gmail.com}
\affiliation[iitb]{Department of Physics, Indian Institute of Technology Bombay, Mumbai 400076, India.}

\author{Ihsan Boustani}
\email{ifboustani@gmail.com}
\affiliation[germ]{Theoretical and Computational Chemistry,
Faculty of Mathematics and Natural Sciences,
Bergische Universität Wuppertal, D-42097 Wuppertal, Germany.}

\author{Alok Shukla}
\email{shukla@iitb.ac.in}
\affiliation[iitb]{Department of Physics, Indian Institute of Technology Bombay, Mumbai 400076, India.}

\title[An \textsf{achemso} demo]
  {Geometry, electronic structure, and optical properties of
boron cages: A first-principles DFT study}
\usepackage{caption}
\usepackage{subcaption}
\abbreviations{IR,NMR,UV}

\begin{document}

\begin{tocentry}
 \begin{center}
\includegraphics[height=4.5cm, width=6cm]{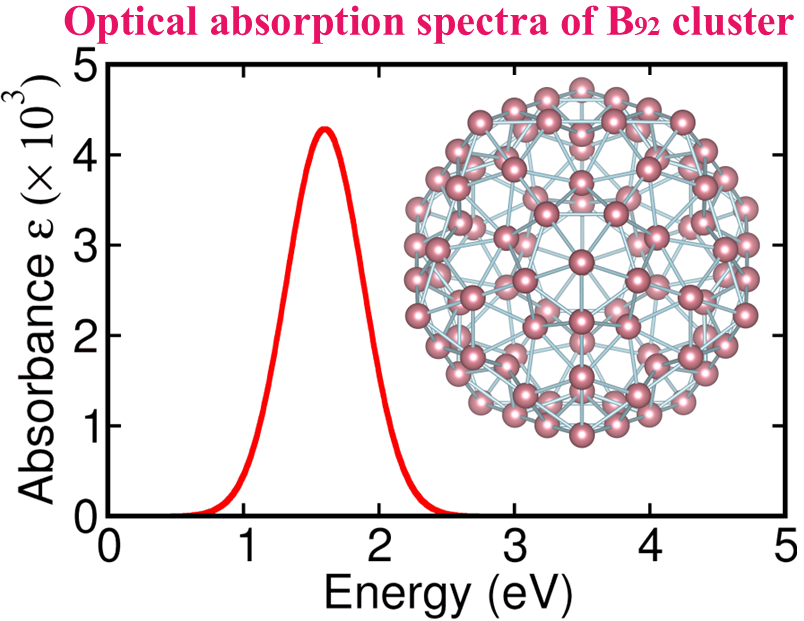}
    \end{center}
\end{tocentry}

\begin{abstract}
A systematic study of the structural, electronic, and optical properties of cage-like boron clusters, with the number of constituent atoms ranging from 20 to 122, has been carried out within the framework of density-functional theory (DFT), employing 6-31G(d, p) extended basis set. The dynamic stability of the clusters is analyzed through the vibrational frequency analysis, while to study the thermodynamic stability, we computed their binding energies per atom. The results suggest that the 32- and 92-atom cages are the most stable among the small and the large structures. The optical absorption spectra of these cages is computed using the time-dependent density-functional theory (TDDFT), which suggests their applications in optoelectronic devices in the visible range of the spectrum. 
\end{abstract}

\section{Introduction}
In the era of nanotechnology, researchers continue to seek increasingly
efficient materials and their nano-structural forms for the technological
applications\cite{zhang2024single}. Among the various nano-structural
forms, cage-like atomic clusters are particularly notable due to their
unique properties and tunability which suggest that they could play
a crucial role in the next-generation devices.
\noindent Boron, due to its neutron-absorbing properties, is a key element with
applications in cancer treatment and nuclear reactors. The carbides,
hydrides, oxides, and nitrides of boron have a wide range of applications
such as hydrogen storage, high energy fuels, anticorrosive coatings,
photocatalysts, etc\cite{kozien2021boron,hagemann2021boron,han2023preparation,verma2023hexagonal,hayat2022fabrication,revabhai2023progress}.
Boron nanostructures have wide-ranging applications in a variety of
fields such as drug design\cite{lesnikowski2016challenges}, and hydrogen
storage \cite{zhao2008boron,wu2009hydrogen,li2009coated,huang2020boron}.
Boron, as a neighbor of carbon in the periodic table with single electron
in the \textit{p} orbital, draws the attention on its nanostructures.
Just as graphene evolved from fullerenes to nanotubes to single atomic
sheets, there is high interest in the potential unfolding of boron,
particularly following the prediction of the B\textsubscript{80}
cage cluster\cite{gonzalez2007b,mishra2025borophene,li2021synthesis}.
The smaller-sized boron clusters tend to acquire a planer geometry,
while the larger ones stabilize in spherical structures \cite{alexandrova2006all,huang2010concentric,zubarev2007comprehensive,mukhopadhyay2009novel}.
Among the spherical cage structures, the core shell type cages where
B\textsubscript{12} is the core cluster, are predicted to be thermodynamically
more stable than the hollow ones\cite{zhao2010b80}. These cage clusters
constitutes as a super-atom in the self-assembled nanostructures\cite{liu2008structural,yan2008face}.
The properties of boron cage-like cluster and their self-assemblies
can be tuned to the desired extent by exohedral and endohedral doping
of impurity atom, \cite{jin2009endohedral,li2009ni,yan2009face}
for example, a magnetic impurity atom that results in the large spin
polarization which could be utilized for the spintronics\cite{chavan2023tunnel}.
The boron compounds and its nanostructures have the excellent optical
properties\cite{tian2019inorganic,revabhai2023progress,yin2016boron,moon2023hexagonal,uddin2023graphene},
however, its atomic clusters are yet to adequately explored for their
optical properties \cite{bhattacharyya2019first}. The the first principles
method, the time dependent density functional theory (TDDFT) within
a accurate basis set provides the description of excited states of
the nanostructures that can be compared with the experiments. On this
ground, it would be valuable to systematically investigate the optical
properties of boron clusters using accurate basis sets, in light of
their potential applicability in the optoelectronic devices.\\
In the present study, we investigate cage-like boron cluster with
the number of constituent atoms ranging from 20 to 122, and discuss
their geometry, stability, electronic structure, and the optical properties.

\section{Computational details}

The calculations for geometry optimization and the optical absorption
spectra for the cage-like atomic clusters of the boron are carried
out using the Gaussian16 code\cite{frisch2016gaussian}, and 6-31G(d,
p) extended basis set. The clusters, along with their isomers, contain
20, 32, 42, 60, 72, 80, 92, 100, 110, and 122 boron atoms. The Benny
algorithm was used for geometry optimization and the convergence thresholds
for maximum force on an atom, and the RMS force were set at 0.023
eV/Å, and 0.015 eV/Å, respectively. The relaxed cages exhibit a variety
of structures such as dodecahedron (B\textsubscript{20}), icosahedron
(B\textsubscript{92}, B\textsubscript{110}), convex (B\textsubscript{80},
B\textsubscript{92}), d5h (B\textsubscript{20}), fullerene (B\textsubscript{80}),
etc. For the relaxed structures, the optical absorption spectra (OAS)
has been studied within the TDDFT as implemented in the Gaussian code\cite{frisch2016gaussian}.
The B3LYP as hybrid exchange-correlation functional and the 6-31G(d,
p) as the localized basis set have been used for this purpose. For
the self-consistent-field calculations, the energy convergence criterion
is set to 0.0002 eV. The GaussView utility is used for the extraction
the output data\cite{dennington2016gaussview}. Unless mentioned otherwise,
the structures studied are charge neutral and in spin singlet state,
and the OAS is calculated for the first 60 excited states in all the
cases. To analyze the thermodynamic stability of these clusters, we
compute their binding energy per atom (B.E.) using the formula.
\[
B.E.=\frac{N\times E(B)-E(B_{N})}{N},
\]
where $N$ is the number of boron atoms in the cage, $E(B)$ is the
total energy of an isolated boron atom, and $E(B_{N})$ is the total
energy of the cage. The total energy of the isolated boron atom obtained
from the SCF calculations is -670.88 eV, used for the binding energy
calculations.

\section{Results and discussion}

\begin{figure}
    \centering
    \includegraphics[width=1.0\linewidth]{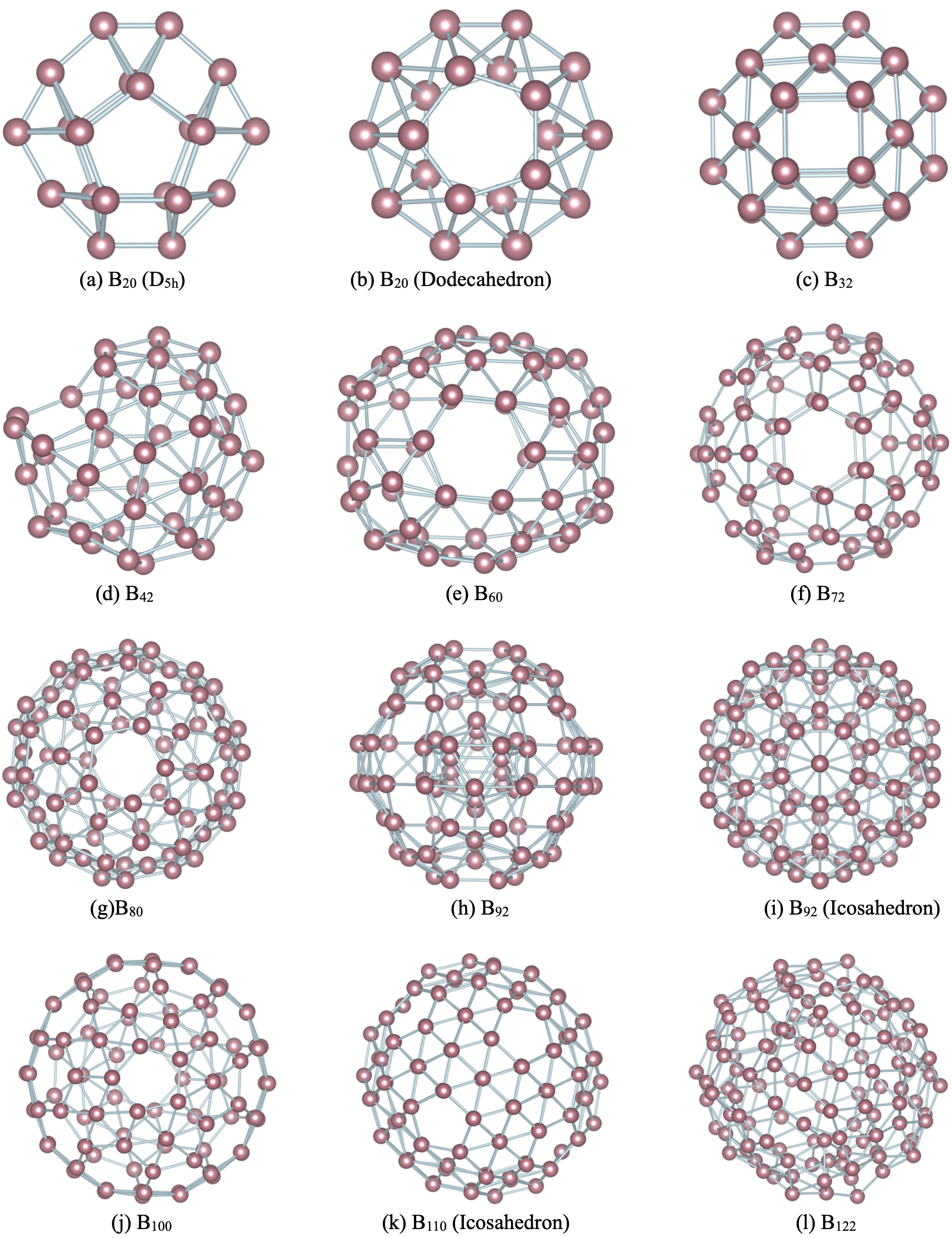}
    \caption{The optimized geometries of the boron cage-like clusters, B\protect\textsubscript{N },
where N denotes the number of constituting boron atoms. The structures
are either hollow or encapsulated cages. }
    \label{GEO}
\end{figure}

To obtain the stable structure for the given configuration, the initial
geometries of clusters are relaxed to local minima on the potential
energy surface. The final optimized geometries of boron cage clusters
are shown in the Figure \ref{GEO}. It is observed in most of the cases that
the clusters retain their cage-like geometry with the exception of
the B\textsubscript{42} cluster, which gets distorted.\\
The B\textsubscript{20} cage has two isomers, the first one with
D\textsubscript{5h} symmetry, while the second one with the dodecahedral
symmetry. In D\textsubscript{5h }case, the structure has overlapping
pentagons with non-uniform inter-pentagon separations, whereas, dodecahedron
has a ring-like top view due to non-overlapping pentagons with a uniform
inter-pentagon separation of 2.65 Å. The intra-pentagon B-B bond length
is 1.66 Å. The sizes of the clusters in D\textsubscript{5h} and dodecahedron
isomer are 5.47 Å and 4.89 Å, respectively. However, their bond-length
range (1.62-1.83 Å) and binding energies (4.69, 4.66 eV) are substantially
close. In the electronic structure, the HOMO-LUMO (H-L) gap is larger
in dodecahedron (2.47 eV) than that in D\textsubscript{5h} (1.51
eV). However, both of these isomers are dynamically unstable as learned
from the vibrational frequencies.\\
The B\textsubscript{32} is the cluster with the largest H-L gap (2.44
eV) that is dynamically stable. This is qualitatively consistent with
the earlier report\cite{sheng2009boron}. The B\textsubscript{32}
structure was created by placing the boron atoms at the center of
each 12 pentagons and 20 hexagons of C\textsubscript{60}. The size
of the optimized structure is 5.44 Å, and its bond length varies between
1.63 Å to 1.78 Å. The final B\textsubscript{32} cluster has eight
hexagons with atoms at their center and six squares. The binding energy per atom of the B\textsubscript{32} cluster is
4.83 eV.\textit{ }Based on its symmetry, H-L gap, and binding energies,
this cluster could be explored for the optoelectronic device applications.

\vspace{-0.1cm}

\noindent The B\textsubscript{42} cluster does not retain its initial structural
symmetry upon relaxation and undergoes distortions, however, the final
structure has some hollow space within it, and can be identified as
a cage. The per atom binding energy of this structure is 5.0 eV, and
the H-L gap is 2.0 eV. Its bond lengths lie in the range of 1.54 Å-1.73
Å with cluster size, 6.75 Å. \\
The B\textsubscript{60} cluster is elongated sphere and that looks
like a barrel-shape structure with septagon at its openings on each
side. The size of the cluster is 8.41 Å, and its per atom binding
energy is 5.15 eV which happens to be the largest among the dynamically
stable clusters. The H-L gap is 1.57 eV and the bond lengths lie in
the range 1.58 Å-1.77 Å. \\
The boron clusters B\textsubscript{20}, B\textsubscript{32}, B\textsubscript{42},
B\textsubscript{60}, atoms have different shapes but for the B\textsubscript{72}
and beyond, the clusters stabilize in the spherical shape. Their surfaces
are combinations of pentagons and hexagons with and without a atom
at their centre.\\
The B\textsubscript{72} cluster comprises of pentagons, hexagons
such that the five hexagons around each pentagon shares a side. Pentagons
have atoms at the center, which are the vertices of this spherical-shaped
cage, whereas the hexagons are planer. The B\textsubscript{72} has
a diameter of 9.0 Å, and its bond lengths lie in the range of 1.60
Å-1.84 Å. The per atom binding energy is 5.10 eV, and the energy gap
is 0.72 eV. Moreover, this cluster is dynamically stable. \\
In B\textsubscript{80}, the structure is similar to that of B\textsubscript{72},
except that pentagons are hollow, and hexagons have atoms at the center.
Its diameter is 8.48 Å, and bond lengths in the range 1.66 Å-1.74
Å. B\textsubscript{80} has a per atom binding energy of 5.17 eV and
an H-L gap of 1.94 eV. However, this B\textsubscript{80} cluster
is dynamically unstable because during the vibrational frequency analysis
we obtained seven imaginary frequencies.\\
In B\textsubscript{92}, two isomers have been studied, convex and
icosahedron. The convex structure is a B\textsubscript{80} cluster
embedded with a B\textsubscript{12} cluster. We note that the six
surface atoms which were at the centre of hexagons, have moved inward
forming bonds with the embedded cluster. The icosahedron comprises
triangles that host the hexagons on the surface; the center atom in
all hexagons has moved inward. The convex structure has a size of
8.82 Å, while the icosahedron has 9.67 Å. The icosahedral structure
is dynamically stable, with an almost double H-L gap (2.07 eV) compared
to the convex case. The range of bond lengths for the convex case
is 1.67 Å-1.84 Å, and for icosahedron, it is 1.63 Å-1.79 Å. The per
atom binding energy of the convex structure (5.15 eV) is slightly
higher than that of the icosahedron (5.10 eV). \\
B\textsubscript{100} cage has twelve pentagons, ten hexagons in hollow
form and eighteen hexagons with an atom at their center. It can be
viewed as three continuous strips of hexagons separated by hollow
pentagons and hexagons. It has hollow pentagons at both poles. The
cage has the most significant per atom binding energy, i.e., 5.18
eV, the smallest H-L gap, 0.60 eV, and is dynamically unstable as
it has one vibrational mode with imaginary frequency. The size of
this cage is 9.74 Å, and bond lengths lies in the range of 1.65 Å-1.77
Å. \\
In B\textsubscript{110} convex structure, the hollow pentagon is
surrounded by five hexagons with atoms at their center. Its size is
9.89 Å and the bond length range is 1.67 Å-1.76 Å. The per atom binding
energy of this cage is 5.10 eV with an H-L energy gap of 0.76 eV.
\\
The B\textsubscript{122} is the largest cluster studied in this report,
with a size extending up to 10.95 Å. It's per atom binding energy
is 5.12 eV, and the H-L gap is 0.89 eV. The bond lengths lie in the
range 1.58 Å- 1.84 Å, and the structure is dynamically stable. The
energetics and structural details of these clusters are given in Table
1. 
\begin{table}[H]
\centering
\begin{tabular}{|c|c|c|c|c|c|c|}
\hline 
\multirow{3}{*}{\textbf{Cluster}} & \textbf{Size } & \textbf{Bond} & \textbf{H-L} & \textbf{Optical} & \textbf{Binding} & \textbf{Dynamic }\tabularnewline
 & \textbf{(Å)} & \textbf{length} & \textbf{gap} & \textbf{gap} & \textbf{energy} & \textbf{Stability}\tabularnewline
 & \textbf{} & \textbf{range (Å)} & \textbf{(eV)} & \textbf{(eV)} & \textbf{(eV)/Atom} & \textbf{}\tabularnewline
\hline 
\hline 
B\textsubscript{20} & 5.47 & 1.6-1.8 & 1.51 & 3.50 & 4.69 & N\tabularnewline
\hline 
B\textsubscript{20} (Dodecahedron) & 4.89 & 1.66-1.83 & 2.47 & 2.29, 3.79 & 4.66 & N\tabularnewline
\hline 
B\textsubscript{32} & 5.44 & 1.63-1.78 & 2.44 & 2.59 & 4.83 & Y\tabularnewline
\hline 
B\textsubscript{42} & 6.75 & 1.54-1.73 & 2.00 & 2.60, 3.38 & 5.01 & Y\tabularnewline
\hline 
B\textsubscript{60} & 8.41 & 1.58-1.77 & 1.57 & 2.41 & 5.15 & Y\tabularnewline
\hline 
B\textsubscript{72} & 9.0 & 1.60-1.84 & 0.72 & 1.43 & 5.10 & Y\tabularnewline
\hline 
B\textsubscript{80}  & 8.48 & 1.66-1.74 & 1.94 & 1.99 & 5.17  & N\tabularnewline
\hline 
B\textsubscript{92}  & 8.82 & 1.67-1.84 & 0.99 & 0.90, 1.80 & 5.15 & N\tabularnewline
\hline 
B\textsubscript{92} (Icosahedron) & 9.67 & 1.63-1.79 & 2.07 & 1.60 & 5.10  & Y\tabularnewline
\hline 
B\textsubscript{100} & 9.74 & 1.65-1.77 & 0.60 & 1.10 & 5.18 & N\tabularnewline
\hline 
B\textsubscript{110}  & 9.89 & 1.67-1.76 & 0.76 & 1.15 & 5.10  & N\tabularnewline
\hline 
B\textsubscript{122} & 10.95 & 1.58-1.84 & 0.89 & 1.29 & 5.12 & Y\tabularnewline
\hline 
\end{tabular}\caption{The size (diameter), range of bond-lengths, HOMO-LUMO (H-L) gap, optical
gap, per atom binding energies, and the dynamic stability of boron
clusters. The multiple optical gaps refers to the shoulder and the
principle peak in the optical absorption spectra. The dynamic stability
based on vibrational analysis is termed as stable (Y) or unstable
(N). }
\end{table}

 \begin{figure}
     \centering
     \includegraphics[width=1.0\linewidth]{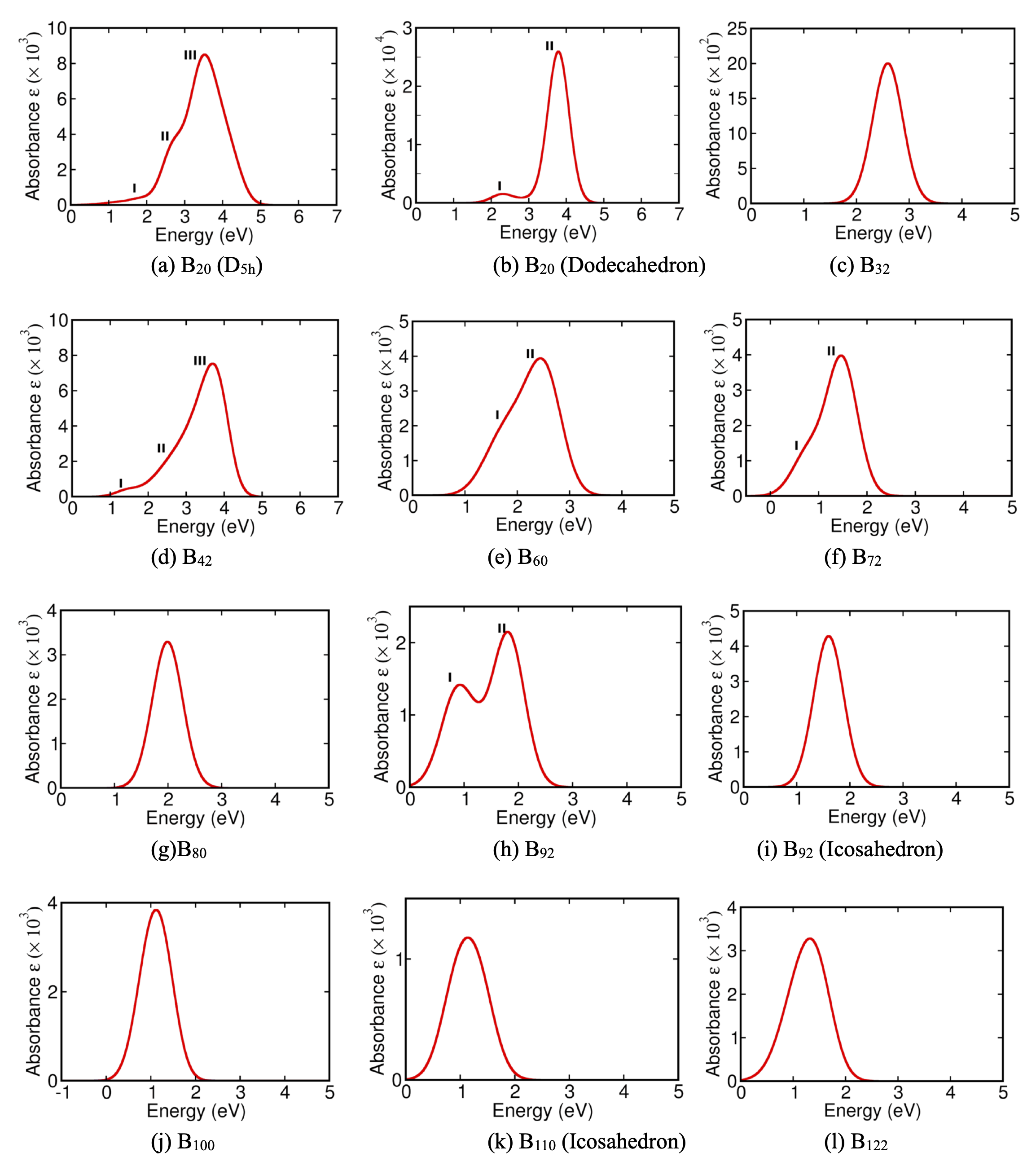}
     \caption{The linear optical absorption spectra of boron clusters computed using
TDDFT approach, at the 6-31G(d,p)/B3LYP level of theory.}
     \label{OAS}
 \end{figure}

\noindent Figure \ref{OAS} shows the OAS for the above discussed boron clusters. Significant
absorption is observed for the incident photon energy up to 7.0 eV.
Often it is seen that the multiple excitations are responsible for
the peaks in the OAS. However, we have considered and listed here
those with significant oscillator strengths (> 0.002) and largest
coefficient for a given energy.\\
The B\textsubscript{20} cluster's OAS has contributions from the
three energy regions. The first peak is a minor shoulder around 1.8
eV, which is due to |H-2 \textrightarrow{} L\textrangle{} excitation,
where H (L) stands for HOMO (LUMO). The second peak is the moderate
shoulder observed around 2.7 eV and has maximum contribution from
the excitations, |H-2 \textrightarrow{} L+1\textrangle{} and |H-6
\textrightarrow{} L+1\textrangle . The third one, which is the principal
peak, occurs around 3.5 eV, with multiple electronic excitations such
as |H-1 \textrightarrow{} L+2\textrangle , |H-8 \textrightarrow{}
L+1\textrangle , |H \textrightarrow{} L+5\textrangle , |H \textrightarrow{}
L+5\textrangle , |H-3 \textrightarrow{} L+3\textrangle , and |H-4
\textrightarrow{} L+3\textrangle{} contributing to it. The other isomer
of B\textsubscript{20} studied has a dodecahedron structure.\\
Among the B\textsubscript{20} cages, the amplitude of optical absorption
for dodecahedron case is higher by an order of magnitude. It has two
peaks in its optical absorption, a minor shoulder peak at 2.3 eV and
a quite intense one at energy 3.8 eV. The excitations corresponding
to the small peak are |H-4 \textrightarrow{} L\textrangle , |H-4 \textrightarrow{}
L+1\textrangle , whereas for the intense peak, that are |H-1 \textrightarrow{}
L+3\textrangle , |H-3 \textrightarrow{} L+4\textrangle , and |H \textrightarrow{}
L+5\textrangle . \\
The B\textsubscript{32} cluster has a single peak at 2.59 eV in its
absorption spectrum with the least intensity among all the clusters.
The excitations correspond to this peak are |H-1 \textrightarrow{}
L+3\textrangle , |H-2 \textrightarrow{} L+3\textrangle , and |H-2
\textrightarrow{} L+4\textrangle{} and these excitations have equal
oscillator strengths. The B\textsubscript{42 }cluster, with a distorted
structure, has two shoulders (I and II) located at 1.2 eV and 2.4
eV, and a major peak (III) at 3.8 eV in its OAS, and the excitations
contributing to them are listed in the Supporting Information (SI).
The OAS of B\textsubscript{60} has a shoulder (I) at 1.8 eV, and
an intense peak (II) at 2.6 eV. The excitations contributing to I
are |H-2 \textrightarrow{} L\textrangle , |H-4 \textrightarrow{} L\textrangle ,
|H-2 \textrightarrow{} L+2\textrangle , |H-2 \textrightarrow{} L+3\textrangle ,
|H-4 \textrightarrow{} L+2\textrangle , and |H-4 \textrightarrow{}
L+3\textrangle , whereas, those to II are |H \textrightarrow{} L+5\textrangle ,
|H-7 \textrightarrow{} L\textrangle , |H-1 \textrightarrow{} L+3\textrangle ,
|H-2 \textrightarrow{} L+4\textrangle , |H \textrightarrow{} L+6\textrangle ,
|H-3 \textrightarrow{} L+3\textrangle , |H-7 \textrightarrow{} L+1\textrangle ,
|H-3 \textrightarrow{} L+5\textrangle , |H-1 \textrightarrow{} L+6\textrangle ,
|H \textrightarrow{} L+8\textrangle , |H-3 \textrightarrow{} L+6\textrangle ,
|H-10 \textrightarrow{} L\textrangle , and |H-10 \textrightarrow{}
L\textrangle .\\
 In B\textsubscript{72}, the optical absorption spectrum is similar
to B\textsubscript{60}, except the absorption edge of B\textsubscript{72}
is at a relatively lower energy. Its first peak which is shoulder
located 0.79 eV is due to the excitations, |H-2 \textrightarrow{}
L\textrangle , |H-2 \textrightarrow{} L+1\textrangle , |H-5 \textrightarrow{}
L\textrangle , |H-4 \textrightarrow{} L+2\textrangle , and |H-5 \textrightarrow{}
L+2\textrangle . The peak is at around 1.54 eV, and the excitations
responsible are |H-3 \textrightarrow{} L+5\textrangle , |H-3 \textrightarrow{}
L+6\textrangle , |H-2 \textrightarrow{} L+4\textrangle , |H-4 \textrightarrow{}
L+3\textrangle , |H-5 \textrightarrow{} L+3\textrangle , |H-4 \textrightarrow{}
L+4\textrangle , |H-5 \textrightarrow{} L+5\textrangle , |H-5 \textrightarrow{}
L+5\textrangle , |H-1 \textrightarrow{} L+7\textrangle , and |H-1
\textrightarrow{} L+10\textrangle . The B\textsubscript{80} cluster
has a single peak in the OAS at energy 1.99 eV due to a single transition;
|H-4 \textrightarrow{} L+3\textrangle . In the B\textsubscript{92}
convex cluster, OAS has two clear peaks at energies 0.8 eV and 1.8
eV. The excitations responsible are |H-3 \textrightarrow{} L\textrangle ,
|H-5 \textrightarrow{} L\textrangle , |H \textrightarrow{} L+1\textrangle ,
and |H-2 \textrightarrow{} L+6\textrangle , |H-3 \textrightarrow{}
L+6\textrangle , |H-4 \textrightarrow{} L+3\textrangle , |H-4 \textrightarrow{}
L+5\textrangle , respectively. In contrast to the B\textsubscript{92}
convex cage, the B\textsubscript{92} icosahedron has only one peak
at 1.6 eV with doubled intensity compared to that of the convex case.
The excitations corresponding to this peak are |H-2 \textrightarrow{}
L\textrangle , |H-1 \textrightarrow{} L+1\textrangle , and |H \textrightarrow{}
L+3\textrangle . The B\textsubscript{100} cluster's OAS also has
a peak around 1.00 eV, and the corresponding transitions are |H-4
\textrightarrow{} L+3\textrangle , |H-4 \textrightarrow{} L+3\textrangle ,
|H \textrightarrow{} L+4\textrangle , |H-1 \textrightarrow{} L+5\textrangle ,
|H-2 \textrightarrow{} L+5\textrangle , and |H-10 \textrightarrow{}
L+3\textrangle . Similarly, B\textsubscript{110} also has a single
peak in the OAS at energy 1.3 eV, and the transitions responsible
for this peak are |H \textrightarrow{} L+3\textrangle , |H \textrightarrow{}
L+4\textrangle , |H-1 \textrightarrow{} L+4\textrangle , |H-1 \textrightarrow{}
L+6\textrangle , |H-1 \textrightarrow{} L+5\textrangle , and |H-1\textrightarrow L+9\textrangle .
In the B\textsubscript{122} case, the peak is around energy 1.3 eV,
to which a large number of excitations listed in the SI contribute.
The excitation energies, their oscillator strengths, and excitations
for the OAS of all the boron clusters are provided in the SI.\\
The binding energies per atom of these clusters are plotted in the
figure 3. It increases from B\textsubscript{12} to B\textsubscript{60},
and beyond that it does not change significantly. For B\textsubscript{12},
it is close to 4.4 eV, and for the constant part, it is around 5.1
eV. 

\begin{figure}
    \centering
    \includegraphics[width=0.7\linewidth]{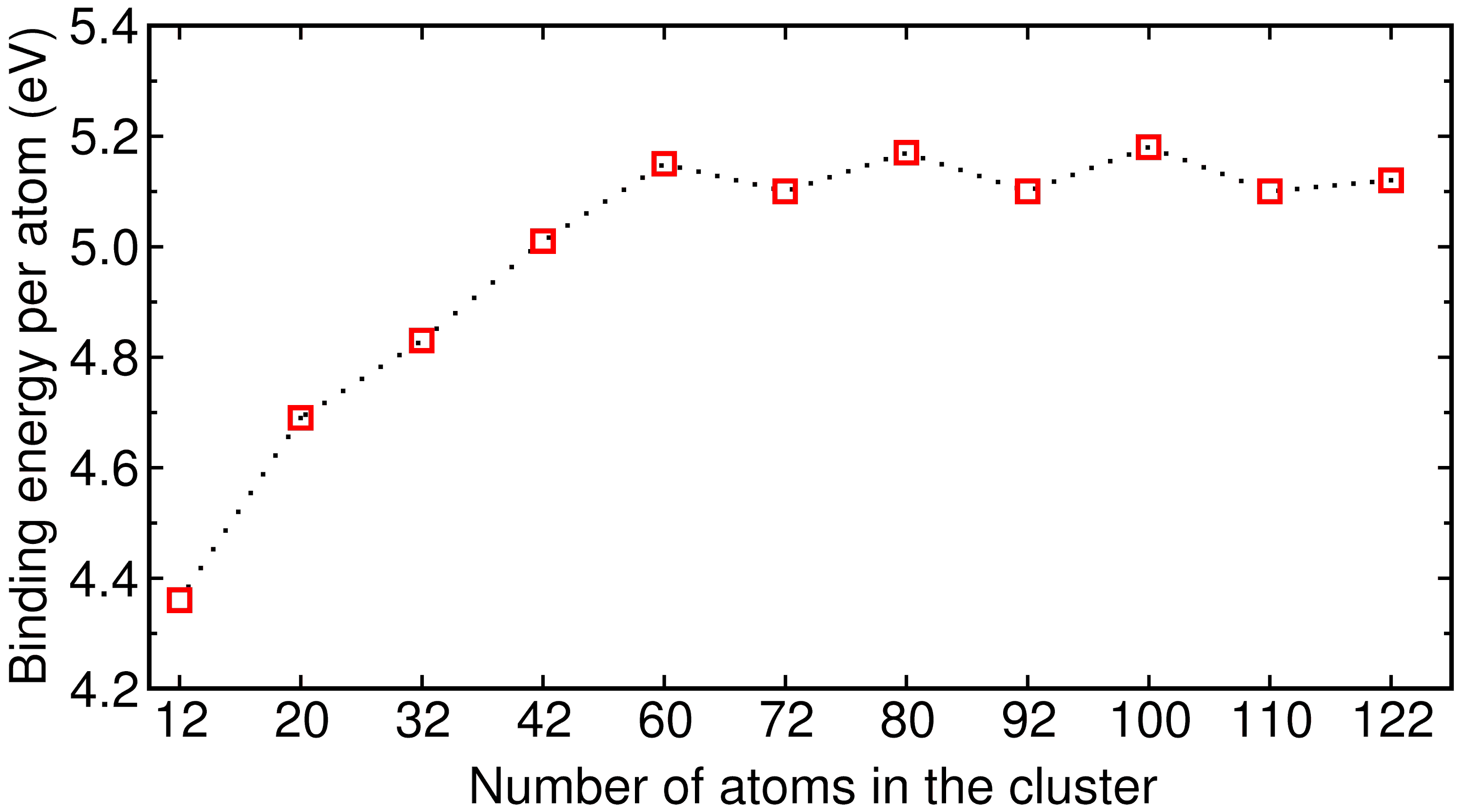}
    \caption{The binding energies per atom of the boron clusters. The dotted line
is a guide to the eye.}
    \label{BE}
\end{figure}

\section{Summary and conclusions}
\noindent Boron cage-like clusters, ranging from 20 atoms with size 4.89 Å to
122 atoms with size 10.95 Å have been studied for structural and optical
properties. The stabilities of these clusters through the vibrational
frequencies and the per atom binding energies are discussed. The per
atom binding energy of the cluster is observed to increase initially
till B\textsubscript{42} atom cluster and beyond that it remains
nearly constant. The clusters B\textsubscript{32}, B\textsubscript{42},
B\textsubscript{60}, B\textsubscript{72}, B\textsubscript{92}(Icosahedron),
and B\textsubscript{122} are found to be dynamically stable. Among
the small clusters B\textsubscript{32} is the most stable one with
the largest HOMO-LUMO gap whereas, among large clusters, B\textsubscript{92}
is the most stable cluster. The optical gap for most of the clusters
lies in the visible spectrum of light which suggests the potential
application of these clusters in the optoelectronics. Since these
clusters provides adequate space inside the cage for encapsulation
or endohedral doping by an atom or molecule, it is of interest to
manipulate the properties of clusters especially in light the optoelectronic
and spintronics device applications. 

\section{Conflicts of interest}

There are no conflicts to declare.

\begin{acknowledgement}
KTC acknowledges Arifa Nazir Bhat and Khushboo Dange, PhD scholars at IIT Bombay for technical discussions.

\end{acknowledgement}




\bibliography{achemso-demo}

\end{document}